# Diffusion pore imaging in the presence of extraporal water

Dominik Ludwig[1,2], Frederik Bernd Laun[3], Karel D. Klika[4], Julian Rauch[1,2,5], Mark Edward Ladd[1,2,6], Peter Bachert[1,2], and Tristan Anselm Kuder[1*]

[1] Department of Medical Physics in Radiology, German Cancer Research Center (DKFZ), Heidelberg, Germany

[2] Faculty of Physics and Astronomy, University of Heidelberg, Heidelberg, Germany

[3] Institute of Radiology, University Hospital Erlangen, Friedrich-Alexander-University Erlangen-Nürnberg, Erlangen, Germany

[4] Molecular structure analysis, German Cancer Research Center (DKFZ), Heidelberg, Germany

[5] Max-Planck-Institute for Nuclear Physics, Heidelberg, Germany

[6] Faculty of Medicine, University of Heidelberg, Heidelberg, Germany

* Email: t.kuder@dkfz.de

## Abstract

Diffusion-weighted imaging (DWI) is a powerful non-invasive tool which is widely used in clinical routine. Mostly, apparent diffusion coefficient maps are acquired, which cannot be directly related to cellular structure.

More recently it was shown that DWI is able to reconstruct pore shapes using a specialized magnetic field gradient scheme so that cell size distributions may be obtained. So far, artificial systems have been used for experimental demonstration without extraporal signal components and relatively low gradient amplitudes.

Therefore, the aim of this study was to investigate the feasibility of diffusion pore imaging in the presence of extraporal fluids and to develop correction methods for effects arising from extraporal signal contributions. Monte Carlo simulations and validation experiments on a 14.1 T spectrometer with a dedicated diffusion probe head were performed.

Both by using a filter gradient approach suppressing extraporal signal components as well as using postprocessing methods relying on the Gaussian phase approximation, it was possible to reconstruct pore space functions in the presence of extraporal fluids with little to no deviations from the expectations. These results may be a significant step towards application of diffusion pore imaging to biological samples. While still ultra-high gradients are required, this problem may be mitigated by recent developments in the field of local gradient coils for human applications.

## 1. Introduction

Diffusion-weighted imaging (DWI) is a powerful and non-invasive tool, which allows gaining insight into the microscopic structure of biological tissues or porous media [1-5]. By using the diffusing water molecules to probe diffusion restrictions, DWI can yield information about structures such as cell membranes on the micrometer scale, which is far below the resolution of MRI in the range of millimeters. Advanced DWI techniques include diffusion tensor imaging, which can be used to reconstruct white matter fiber pathways in the human brain [3, 6, 7]; diffusion kurtosis imaging used to probe diffusion heterogeneity [8-10]; exploration of closed boundaries in the short- and long-time limit [5, 11-13]; or double-diffusion encoded imaging for investigation of cell membrane permeability [14-18]. In clinical routine, DWI is most prominently used for tumor [19-24] and stroke diagnosis [25-27] where the apparent diffusion coefficient (ADC) is measured in most cases.

While the ADC can be used to estimate the typical diffusion distance of water molecules in tissue, the information derived from conventional DWI is in most cases only indirectly connected with the restricting geometries. DWI reaches high sensitivities for detection of malignant lesions; specificity, however, remains often limited [22, 28]. Parameters more directly related to the geometry of diffusion restrictions, such as cell size and shape distributions, could thus result in a better correlation with histology, preventing unnecessary biopsies and allowing for a better sampling of tumor heterogeneity compared to invasive biopsies. However, it has been unknown for a long time whether DWI enables direct access to the shape of closed cells or pores. This problem was solved by Laun *et al.*, who showed that the direct measurement of the shape of arbitrary closed pores filled with an NMR-visible medium is possible using a modified DWI sequence [12]. In contrast to commonly used DWI approaches, a combination of a long magnetic field gradient with low amplitude and a short strong gradient pulse was introduced. This gradient profile conserves the phase information and therefore enables the reconstruction of pore shapes by a simple Fourier transform. Since all pores in the volume of interest contribute to the resulting signal, a significantly higher signal-to-noise ratio and thus resolution can be reached by diffusion pore imaging than by using conventional MR imaging. Subsequently, various pore imaging approaches [29-33] have been introduced and demonstrated experimentally in phantom experiments using capillaries [34-37] and in larger pores with hyperpolarized xenon-129 gas as the MR-detectable medium [13, 32, 33, 38]. So far, to our knowledge, only closed pores without extraporal signal components have been investigated in detail theoretically and experimentally; for example, capillaries dried on the outside were employed [36, 37]. However, for applications to cells and biological tissues, extracellular signal components have to be considered, even if the diffusion times used are short enough to consider cells as closed pore geometries. Furthermore, gradient amplitudes of typically more than 1 T/m are needed to acquire pore images for structures in the range of cell sizes [30], while for the experimental validations usually gradient amplitudes below 1 T/m were used in combination with larger pore sizes.

Therefore, the aim of this work was the investigation of the applicability of diffusion pore imaging in the presence of extraporal water by performing Monte Carlo simulations and measurements using a 14.1 T Bruker spectrometer and dedicated diffusion probe with a maximum gradient amplitude of $G_{max}$ = 18 T/m. In order to mitigate distorting effects from extraporal signal contributions, correction methods using postprocessing were developed and the original gradient scheme was extended by a filter gradient approach. This approach allows for an effective suppression of effects resulting from extraporal signals, so that direct measurements of pore images in a surrounding MR-detectable medium become possible.

## 2. Theory

In order to show that the pore space function can be directly measured using pore imaging, we reiterate the basics equations presented previously [13, 30, 39] and consider an arbitrary closed pore as shown in Figure 1a. The pore can be characterized by the pore space function $\rho(x)$ which equals 0 outside and $1/V$ inside the pore.

The asymmetric gradient scheme used in this work for diffusion pore imaging (Figure 1b) is called long-narrow approach [12]. In the following, the index "L" denotes the long and "S" the short gradient pulse. The temporal gradient profile is defined by

$$G_{\delta_S,\delta_L}(t) = \begin{cases} G_L = -\frac{\delta_S}{\delta_L} G_S & \text{for } 0 \leq t \leq \delta_L \\ G_S & \text{for } \delta_L \leq t \leq T \\ 0 & \text{else,} \end{cases} \quad (1)$$

where $T = \delta_L + \delta_S$ is the total length of the gradient scheme. The rephrasing condition $\int_0^T G_{\delta_S,\delta_L}(t)dt = 0$ is fulfilled.

The applied $q$ value for the gradients can be calculated according to:

$$q = \gamma G_S \delta_S = -\gamma G_L \delta_L \quad (2)$$

In general, the diffusion-induced signal attenuation can be calculated using

$$S_{\delta_L,\delta_S}(q) = \langle e^{-i\gamma \int_0^T G_{\delta_L,\delta_S}(t) \cdot x(t)dt} \rangle, \quad (3)$$

where $\langle \cdot \rangle$ indicates the average over all possible random trajectories of the particles. $S_{\delta_L,\delta_S}(q)$ can be represented in terms of the center of masses of the trajectories during the long and short gradient pulses [12, 40]:

$$S_{\delta_L,\delta_S}(q) = \langle e^{iq \cdot (x_{cm,L} - x_{cm,S})} \rangle, \quad (4)$$

where the center of masses can be calculated according to:

$$x_{cm,L} = \frac{1}{\delta_L} \int_0^{\delta_L} x(t)dt \quad (5)$$

$$x_{cm,S} = \frac{1}{\delta_S} \int_{t=T-\delta_L}^{T} x(t)dt. \quad (6)$$

For the case of $\delta_L \to \infty$ and $\delta_S \to 0$ while keeping $q$ constant, a random walker will explore the entire pore during the long gradient; therefore, the center of mass of the "long" trajectory will converge to the pore center of mass $x_{cm}$. For an infinitely short second gradient pulse, the center of mass of the "short" trajectory will become equal to the trajectory endpoint $x_2$. The signal for $\delta_L \to \infty$ and $\delta_S \to 0$ can then be calculated as follows:

$$S_{\infty,0}(q) = e^{iq \cdot x_{cm}} \langle e^{-iq \cdot x_2} \rangle = \int \rho(x_2) e^{-iq \cdot x_2} dx_2 = e^{iq \cdot x_{cm}} \tilde{\rho}(q) \quad (7)$$

If the long-time limit is fully reached ($T \to \infty$), the long gradient does not lead to any signal loss, but only to an additional phase proportional to $x_{cm}$, which is identical for all spin packets. Thus, Eq. (7) contains phase information allowing for a direct reconstruction of the pore shape from the measured signal $S_{\infty,0}(q)$ by applying an inverse Fourier transform ($iFT$). For the case of multiple pores inside a given voxel, the measured pore space function is given by [39]

$$\rho_{\text{avg}}(x) = iFT\left(S_{\infty,0}(q)\right)(x) = \sum_{n=1}^{M} f_n \rho_{n,0}(x) \tag{8}$$

where $f_n = V_n/V$ is the volume fraction of the $n$-th pore of the total volume $V$ of the NMR visible medium within all pores and $\rho_{n,0}(x) = \rho(x + x_{n,\text{cm}})$ the pore space function of the $n$-th pore with its center of mass $x_{n,\text{cm}}$ shifted to the origin. Therefore, in the case of multiple pores, diffusion pore imaging leads to the measurement of the arithmetically averaged pore space function. This also means that all pores inside a given volume contribute to the signal and therefore allow for a much higher resolutions compared conventional MRI, which is highly SNR-limited when enhancing resolution.

When considering the influence of extraporal signal components, an important parameter is the packing density which is quantified by the extraporal volume fraction

$$f_e = \frac{V_{\text{extraporal}}}{V_{\text{total}}}, \tag{9}$$

yielding the quotient of the volume of the NMR visible medium in the extraporal space divided by the total volume of the diffusing medium.

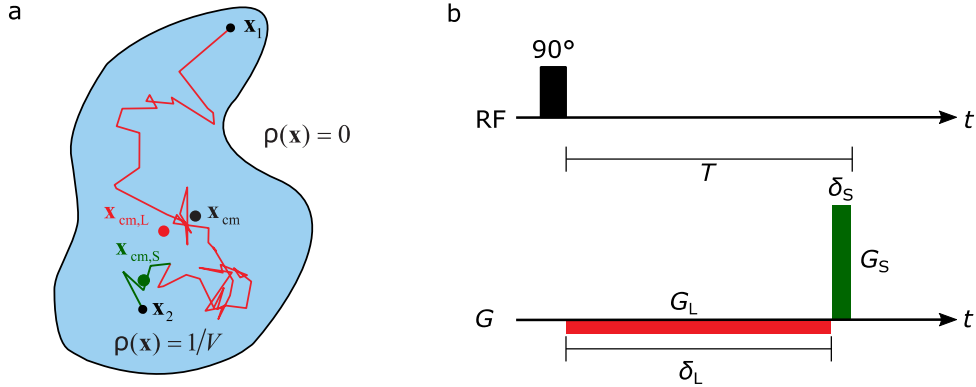

Figure 1: (a) closed pore with trajectories of a random walker during the gradient pulses shown in (b). The center of mass of the trajectory during the long gradient (red) $x_{cm,L}$ will converge to the pore center of mass $x_{cm}$ for $T \to \infty$. This will result in an additional phase $exp(iq \cdot x_{cm})$. For $\delta_S \to 0$ for the short gradient, the center of mass $x_{cm,S}$ of the corresponding trajectory (green) will become equal to the trajectory endpoint $x_2$.

## 3. Material and Methods

### 3.1 Removal of extraporal signal contributions

In this work, two approaches are suggested to perform diffusion pore imaging in the presence of extraporal water: Postprocessing methods and the application of additional filter gradients to suppress unwanted signal components. The most important assumption used here for separation of the intra- and extraporal signal fractions is a higher effective diffusion coefficient for the extraporal water for the long diffusion times used for pore imaging experiments.

For the postprocessing methods, it is additionally assumed that the extracellular component follows the Gaussian phase approximation [4, 41, 42], so that the signal can be modeled using $S_{\text{extra}}(q) = S_{0,\text{extra}} e^{-q^2 c_1}$ with a constant $c_1$ for radial acquisitions in $q$ space. Due to the linearity of the Fourier transform, the extraporal component can be modeled by a Gaussian function $S_{x,\text{extra}}(x) =$ $S_{x,0,\text{extra}} e^{-x^2 c_2}$ in $x$ space as well, so that the total signal obtained from radial projections in $q$ space can be written as

$$S(x) = S_0 \left[ f_e \, e^{-x^2 c_2} + (1 - f_e) \, iFT(S_{\delta_L, \delta_S})(x) \right] \quad (10)$$

with the extraporal or extracellular signal fraction $f_e$, which equals the extraporal volume fraction in the case of equal relaxation times. This contribution can then be fitted in $x$ space using the signal region outside the expected pore sizes and thus be retrospectively removed from the measured data. This technique is applicable as long as the signal drop of the extraporal component is significantly faster than that of the intraporal signal in the long-time limit so that $e^{-x^2 c_2}$ becomes broad in $x$ space compared to the extent of the considered pores. Figure 2 sketches this approach for artificial data.

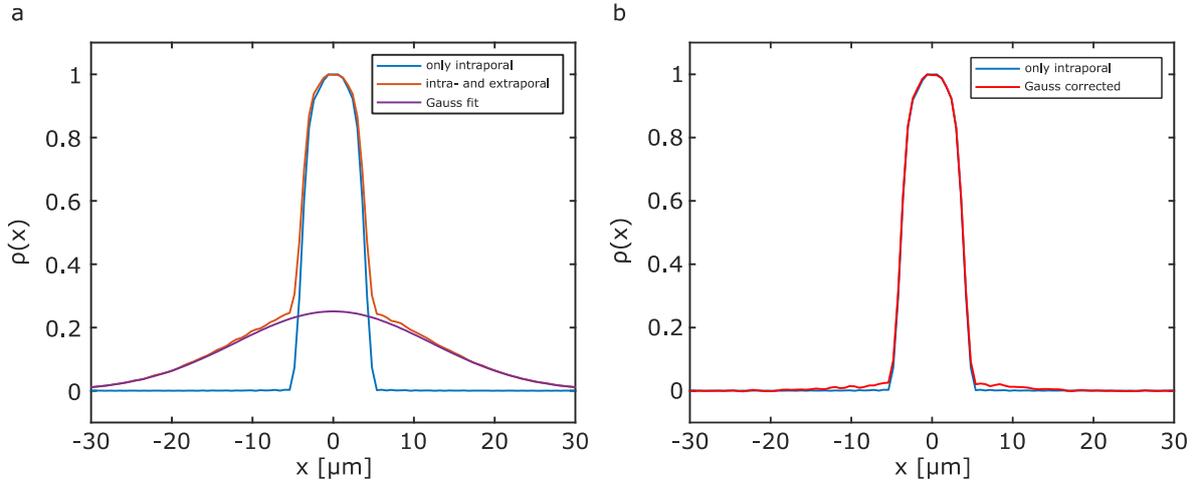

Figure 2: (a) Illustration of the workflow for removal of the extraporal signal based on the Gaussian phase approximation. A Gaussian function (violet) is fitted to the outer part of the pore function and then subtracted from the actual pore space function (brown). (b) The corrected pore space function (red) shows nearly no deviation from the case of pure intraporal signal. All pore space functions are normalized for better comparison.

Again starting from the Gaussian phase approximation, we propose an additional method for retrospectively removing the extraporal signal component. The intraporal signal follows the Gaussian phase approximation for sufficiently low $q$ values as well. We expand this Gaussian phase approximation for small $q$ and assume that the intraporal signal can be described by

$$S_{\text{intra}}(q) = S_{0,\text{intra}} \cdot \left( 1 - \left( \frac{q}{a} \right)^2 \right) \quad (11)$$

with a constant $a$. We further assume that the extraporal signal drops so fast with increasing $q$ that there is a range of $q$ values, where the extraporal signal has already dropped to a negligible signal fraction, but the intraporal signal can still be described by the Gaussian phase approximation (Eq. 11). This assumption may be dependent on the geometry and the experimental parameters. For the cylindrical domain, this is fulfilled near the first zero crossing, as demonstrated below. The long-narrow approach itself leads to a sufficient suppression of extraporal contributions for larger $q$ values ($q > 0.3 \, \mu m^{-1}$) due to the higher effective diffusion coefficient. Using the appropriate range of $q$ values, Eq. (11) is then used to extrapolate the intraporal signal to $q = 0$.

## 3.2 Simulations

Simulations were carried out using an in-house developed Monte-Carlo Simulation Tool, implemented in MATLAB (The MathWorks, Natick, Massachusetts, USA) using CUDA GPU-acceleration. The step size of each random walker was $dr = \sqrt{6\,D\,dt}$ scaled by uniformly distributed and independent random numbers between -1 and 1 for each direction. Random walkers encountering obstructions (e.g. border of the triangles and cylinders) were elastically reflected. A more detailed description regarding the simulation tool and signal calculation can be found in [18]. Two-dimensional pore images were generated by simulation of 32 uniformly distributed gradient directions, equivalent to radial spokes in conventional images. The projection of the pore space function for each spoke was then calculated using an inverse Fourier transform. Two-dimensional pore images were finally reconstructed using an inverse Radon transform applied on the acquired spokes. The simulation parameters are listed in Table 1.

Table 1: Simulation parameters used for this study. $D_i$ and $D_e$ denote the free intra- and extraporal diffusion coefficient. For the definition of the NMR sequence parameters see Fig. 4.

| Edge length triangle [µm] | 10 |
|---|---|
| $d_{cylinder}$ [µm] | 10 |
| $D_i = D_e$ [µm²/ms] | 2 |
| $\delta_L$ triangle [ms] | 197.5 |
| $\delta_S$ triangle [ms] | 2.5 |
| $\delta_L$ cylinder [ms] | 117 |
| $\delta_S$ cylinder [ms] | 2.3 |
| $f_e$ triangle | 0.6 |
| $f_e$ cylinders | 0.3 |
| $G_{max}$ [T/m] | 6.3 |
| #particles | $10^6$ |
| $dt$ [µs] | 4 |
| $T_F$ [ms] | 80 |
| $\Delta_F$ [ms] | 40 |
| $G_F$ [T/m] | 0.03 |
| $b_{filter}$ [s/mm²] | 2750 |

## 3.3 Validation experiments

Validation experiments were carried out using a Bruker 14.1 T spectrometer (Bruker Biospin, Ettlingen Germany) equipped with a Diff30 probe head with a maximum gradient amplitude of $G_{max} = 18$ T/m in z-direction and an 8-bit raster resolution. Due to the limited gradient step resolution leading to rounding errors and the large difference in amplitudes between the short and long gradient pulses, it was necessary to compensate for errors due to the gradient scaling by adjusting the length of the short gradient pulse in order to fulfill the refocusing condition. This was performed using calibration measurements prior to the actual experiments.

Phantoms consisted of fused silica glass capillaries (Molex, Lisle, USA) with inner diameters of $d_{inside} = 10$ µm and 25 µm shielded by polyamide which results in the outer diameter $d_{outside} = 150$ µm. A schematic representation of the capillaries can be found in Figure 3a. The capillaries were cut to a length of 3.5 mm and stacked inside an acrylic cylinder within a 1 mm diameter bore in the short axis of the cylinder. A schematic sketch of the phantom holder is shown in Figure 3b. In total, about 40 capillaries were placed inside the bore. The total water content inside the pores amounted to about 10 µL. The phantom holder with the capillaries was then placed inside a 5 mm NMR-tube (see Fig. 3c) and fixed in place by a Teflon spacer. Due to the perpendicular orientation of the capillaries to the main magnetic field and the different magnetic susceptibilities involved, it was necessary to adjust the susceptibility by using 5 molar NaCl solution in- and outside of the capillaries. Furthermore, the long narrow approach shown in Figure 1b cannot directly be used for measurements, since $T_2^*$ is too short to reach the long-time limit. Therefore the long gradient was split into a CPMG-like gradient train [36] with segments of about 10 ms each including ramptimes of $t_{\mathrm{ramp}} = 0.3$ ms, in order to reach the necessary long-time limit. A schematic representation of this sequence can be found in Figure 3a. All measurements were acquired using 512 averages, 21 $q$ values and a repetition time of $\mathrm{TR} = 5$ s. The complete set of measurement parameters is shown in Table 2. Additionally, the most important parameters are repeated in the results figures.

Table 2: Measurement parameters used in this study

| General parameters | |
|---|---|
| #$q$ values | 21 |
| TR [s] | 5 |
| # averages | 512 |
| Extrapolation measurement | |
| $\delta_l$ [ms] | 10.22 |
| # long segments | 39 |
| $\delta_L$ [ms] | 398.6 |
| $\delta_S$ [ms] | 1.9 |
| $t_{\mathrm{ramp}}$ [ms] | 0.3 |
| $G_{\mathrm{max}}$ [T/m] | 2.7 |
| Gauss correction and filter measurement | |
| $\delta_l$ [ms] | 10.5 |
| # long segments | 11 |
| $\delta_L$ [ms] | 115.5 |
| $\delta_S$ [ms] | 2.3 |
| $t_{\mathrm{ramp}}$ [ms] | 0.3 |
| $G_{\mathrm{max}}$ [T/m] | 6.3 |
| $T_F$ [ms] | 84 |
| $\Delta_F$ [ms] | 42 |
| $b_{\mathrm{filter}}$ [s/mm] | 2750 |

For suppression of extraporal signal contributions an additional diffusion weighting was added in front of the actual long-narrow approach (see Fig. 4b). This so-called filter consists effectively of two long gradient pulses. Ideally, during each of these pulses, the motional narrowing regime is reached so that negligible diffusion-induced signal loss inside the pores is achieved, while the extraporal signal is effectively suppressed. In order to choose the right echo pathway, it was necessary to use an appropriate phase cycling approach for the 180° pulses, which can be found in Appendix Table A1.

For reference measurements without extraporal signal contributions, the capillaries were submersed in Flourinert[TM] (F-770, 3M), which displaces the extraporal water because of its three-times higher density and has no [1]H-detectable MR-signal. The capillaries themselves were again filled with 5 molar NaCl solution.

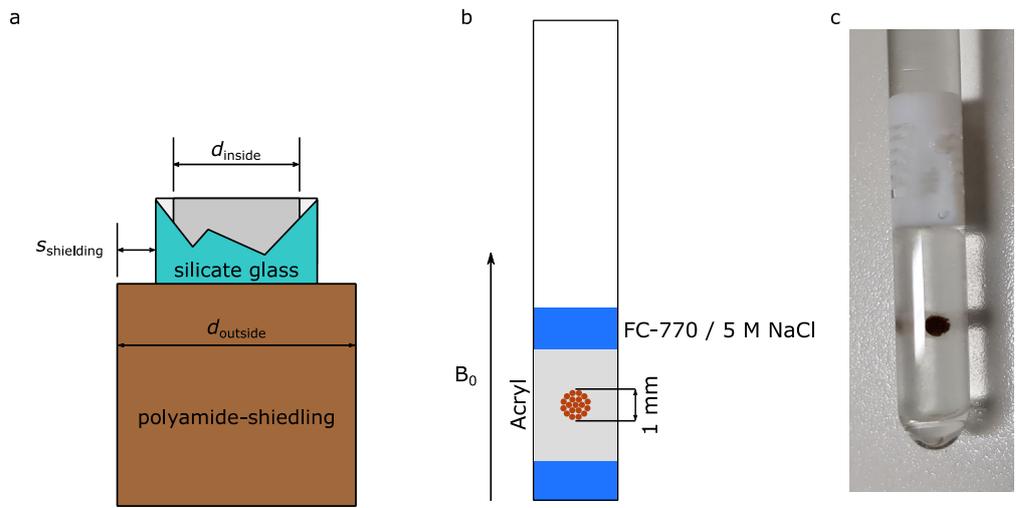

Figure 3: (a) schematic sketch of the acrylic phantom holder inside the NMR-tube with the capillaries placed along the orthogonal direction of the main axis of the cylinder. (b) actual capillary phantom, consisting of 40 capillaries with a length of 3.5 mm placed orthogonal to the main magnetic field, amounting to an intraporal water volume of roughly 10 µL.

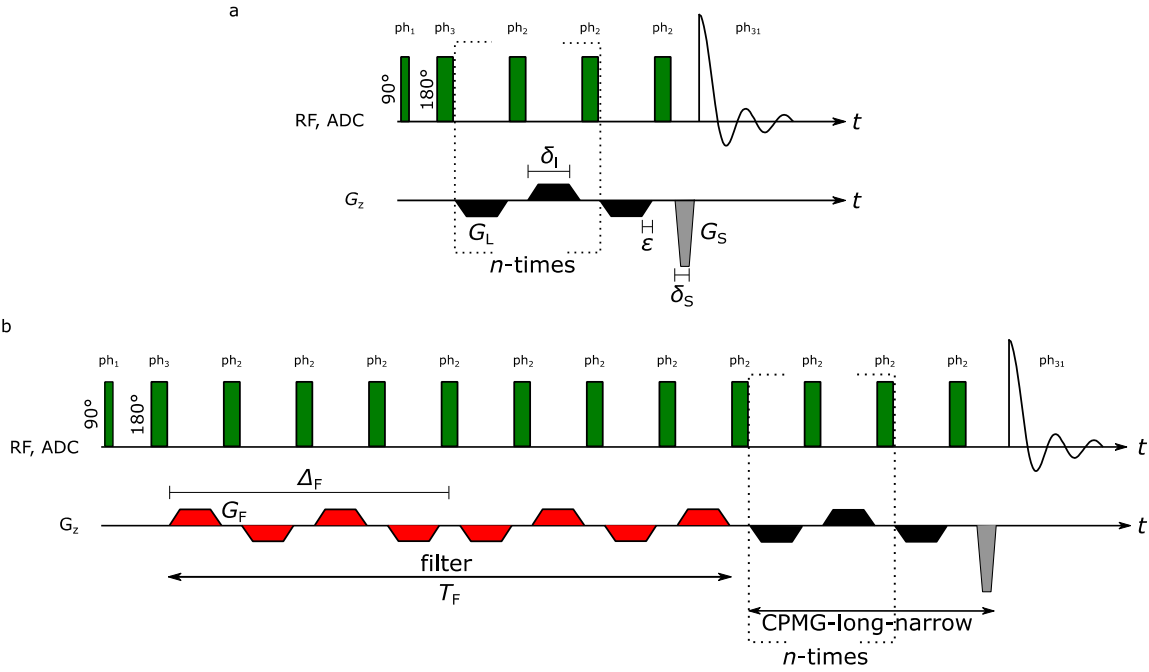

Figure 4: Schematic representation of the CPMG-like long-narrow implementation without (A) and with (B) the proposed filter approach. Splitting of the long gradient pulse was necessary to reduce the effect of static inhomogeneities and thus to maintain a sufficient signal in the long-time limit. The dotted segment is repeated $n$ times to yield the appropriate duration of the long gradient. The filter duration can be adjusted in a similar way.

## 4. Results

### 4.1 Simulation results

#### 4.1.1 Equilateral triangle

Figure 5(a,c) shows the simulated signal using equilateral triangles with an edge length of 10 µm. Signals were simulated using the CPMG-like implementation of the long-narrow approach with and without the filter. The measured pore imaging signal along the exemplary $x$- and $y$-direction is shown in (a) and (b), respectively. The signal along the $x$-direction does not have an imaginary part, while along the y-direction there is an imaginary contribution. The simulated signal with extraporal contributions deviates largely for the first $q$ values from the intraporal one, while for larger $q$ values no deviation from the signal with only intraporal contribution can be observed due to the relatively low packing density and high extraporal effective diffusion coefficient. The signal simulated with the filter approach shows no deviation from the intraporal signal except for $q = 0$ in all directions.

The corresponding pore space functions are shown in (c) and (d). For both directions, the measured signal with extraporal contribution clearly differs from the pore space function obtained for intraporal particles only. When correcting the simulated signal using the Graussian fit in $x$ space, the resulting pore space function is in good agreement with the expectation. The same is valid for the simulation using the additional filter in front of the CPMG-like long-narrow approach, however, a small but negligible baseline remains in the pore space function.

The reconstructed two-dimensional pore images for the equilateral triangle are shown in Figure 5(e-g) for the different acquisition methods. Images were reconstructed using all 32 simulated gradient directions for the long-narrow approach and applying an inverse Radon transform to the projection data. For better comparability, images were normalized to their individual maximum value. While the triangle is still visible in the case of intra- and extraporal signal contributions (e), the contrast is weak and the boundaries of the pore are not clearly visible. For the gauss-corrected simulation (f) as well as the simulation with the filter approach (g) the contrast is significantly increased, and the boundaries are clearly visible. The small baseline in the pore space function of the filter approach also leads to a small baseline in the reconstructed pore image.

#### 4.1.2 Cylinders

Figure 6a shows the simulated signal using cylinders with a diameter of 10 µm. Signals were simulated using the CPMG-like implementation of the long-narrow approach with and without the filter. Since the cylinders are radially symmetrical, only one exemplary gradient direction ($x$ direction) is shown here.

Similar to the equilateral triangles, the simulated signal with extraporal contributions differs significantly for small $q$ values ($q < 0.5/\mu m$) from the signal with only intraporal contributions. The filter method shows no deviation from the expected signal, even for small $q$ values. The extrapolation approach is also displayed in Figure 6a. The fit was performed using the simulated signal with intra- and extraporal contributions for $q$ values close to the first zero-crossing leading to the dashed gray line.

The simulated signal was then replaced by the extrapolation approach ($S_0 \cdot (1 - (q/a)^2)$) for $q$ values up to the first zero-crossing of the simulated NMR signal. The resulting signal is represented as solid gray line. The extrapolated signal deviates quite significantly for the lowest $q$ values due to underestimation of $S_0$ in the fit. The reconstructed projections of the pore space function on the $x$ gradient axis are shown in Figure 5b.

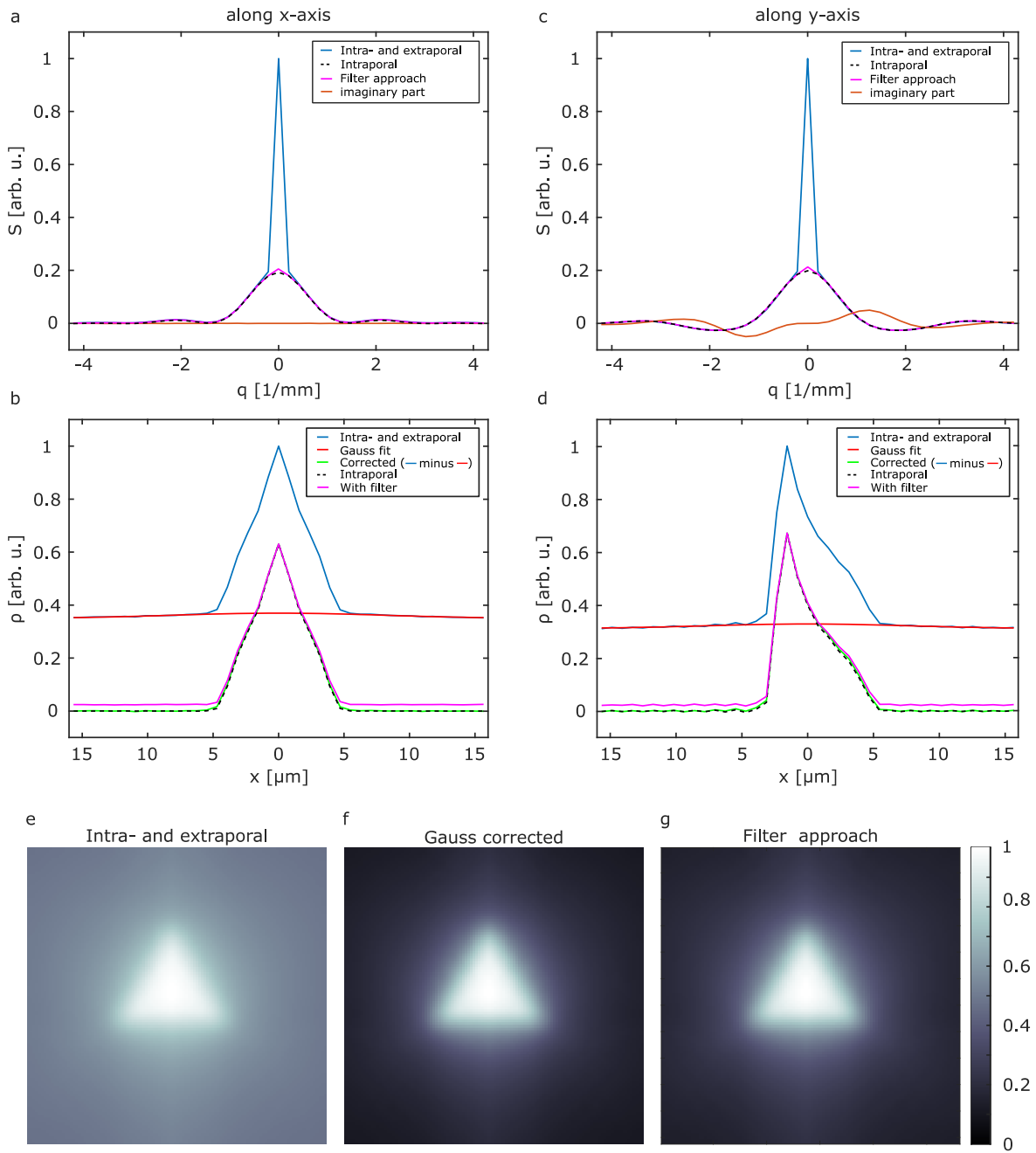

Figure 5: (a,c) Simulated signals for different gradient directions for the equilateral triangles with a extraporal volume fraction of $f_e = 0.6$. The corresponding projection of pore space functions on the respective gradient axis (b,d) for the signals with the different correction methods. (e-g) reconstructed two-dimensional pore images.

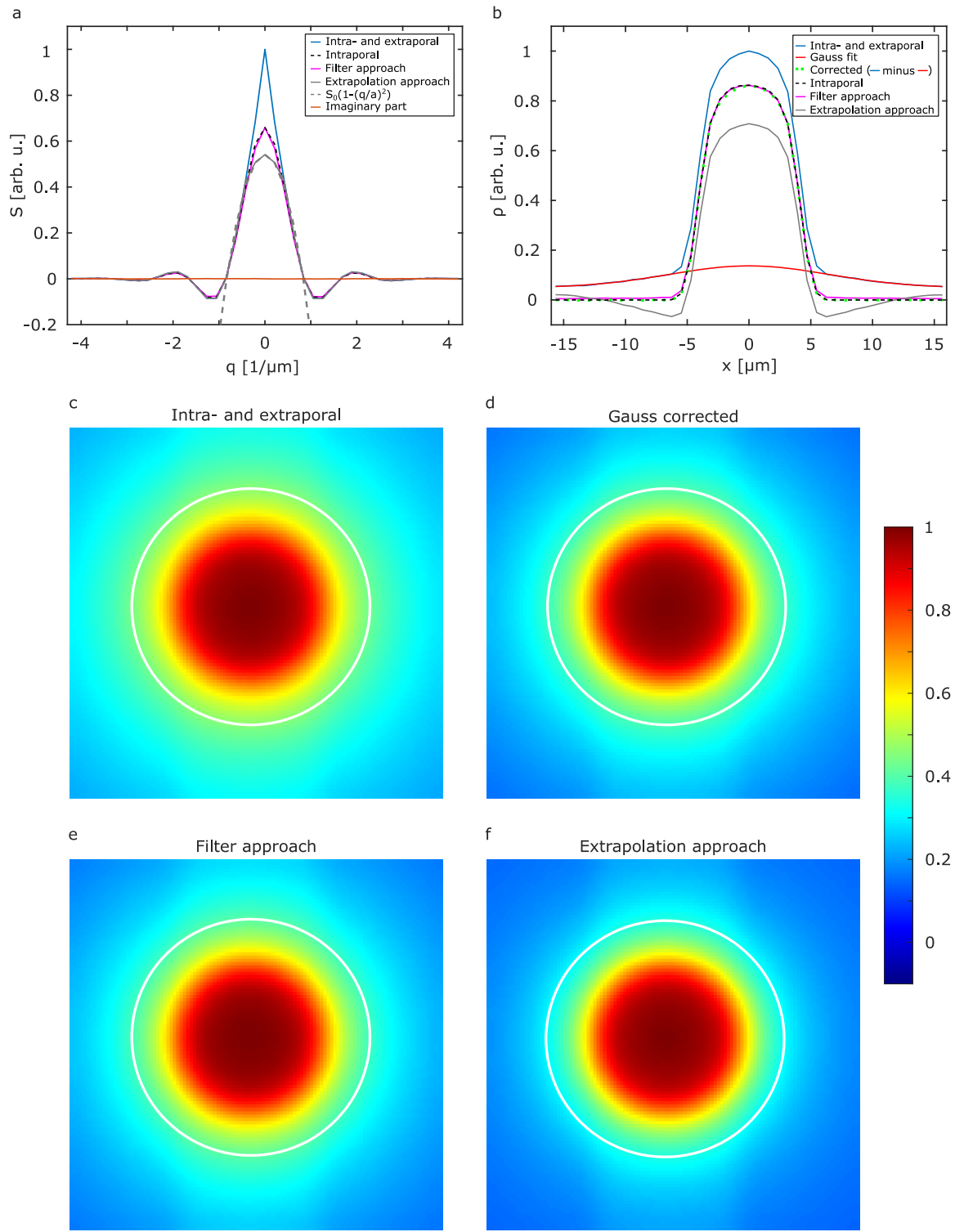

Figure 6: (a) Simulated and corrected signals for the cylinders with a extraporal volume fraction of $f_e = 0.3$. The corresponding projection of the pore space function on the gradient axis (b) for the signals with the different correction methods (extrapolation in $q$ and $x$ space and filter gradients). (c-f) reconstructed two-dimensional pore images. All correction methods lead to an increased contrast.

The pore space function for the simulation with intra- and extraporal contributions again displays a large baseline which can be fitted by a Gaussian for the correction method. The pore space functions simulated using filter gradients and the Gauss correction in $x$ space are in good agreement with the intraporal function, while the extrapolation approach in this case deviates quite significantly from the expected function. However, the pore size is reproduced correctly for all methods. The reconstructed two-dimensional normalized images are shown in Figure 6(c-f). Due to the smaller extraporal fraction $f_e = 0.3$ the differences in the reconstructed two-dimensional pore images of the cylinders are relatively small compared to the case of equilateral triangles in Fig. 4, which is highlighted by a different colormap.

The image with intra- and extraporal contributions (c) shows the largest baseline and therefore has the lowest contrast. All the correction methods yield similar images. While the extrapolation method seems to yield the best result in this case; however looking at the projection data in (b), the extrapolation method performs worst, so that the visual inspection of the 2D images might lead to a biased conclusion.

### 4.2 Measurements

In order to demonstrate the feasibility of the correction methods, spectrometer measurements with cylindrical pores were carried out. Since the Diff30 diffusion probe only generates a gradient along the $z$ axis, the following results show only one-dimensional projections of the pore space functions on the $z$ axis.

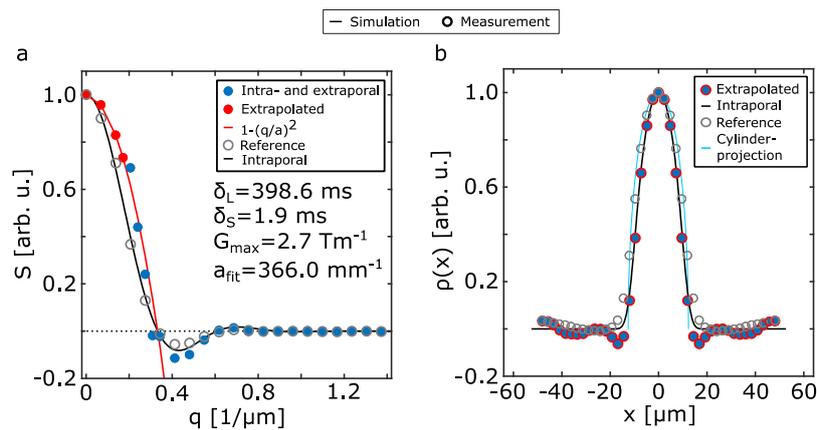

Figure 7: $q$-Space extrapolation approach: (a) Simulated ground truth data with only intraporal contribution (black), as well as measured signals with (blue) and without (gray) extraporal contribution for the glass capillaries (25 µm). Furthermore, the extrapolated signal from signal points close to the zero-crossing is shown in red. The resulting pore space functions are shown in (b). The extrapolation approach is in good agreement with both the simulated ground truth and the measurement with only intraporal contribution. (b) Corresponding projection of the pore space function for the signals with the different correction methods.

In order to yield better results for the extrapolation approach, a denser q-sampling close to the zero-crossing was used. The resulting fit-curve is shown in red. The measurements are shown in Figure 7a.

As a reference, a simulation containing only intraporal signal is shown in black. Additionally, experiments were carried out with Fluorinert[TM] displacing the extraporal signal contributions, referred to as reference measurements in the following (gray circles). The reference measurement is in good agreement with the expected signal, while the extrapolation approach slightly differs. The resulting projections of the pore space functions are shown in Figure 7b. The intraporal reference measurement, the simulation and the pore space function obtained with the extrapolation approach (blue dots with red edges) in $q$ space are in relatively good agreement with each other and the ideal projection of the cylinder geometry (cyan). A Gauß fit in $x$ space was not possible for this measured data, probably due to insufficient signal quality at low $q$ values resulting from insufficient suppression of other coherence pathways leading to deviations from the expected Gaussian signal shape in the outer regions in $x$ space.

### 4.2.2 Gauss-correction and filter approach

Measurements of capillaries with a diameter of 10 µm can be found in Figure 8a. Parameters used for the measurement are stated within the Figure. The measurement with the filter approach is shown in magenta and without in blue. Again, the deviation for the signal for small $q$ values is quite significant. It has to be noted that the packing density was relatively low ($f_e \sim 0.8$) due to the large outer diameter of the fused silica capillaries and the NaCl solution surrounding the cylinders. The measured signal with the filter approach is again in good agreement with the expected signal from intraporal simulations (black dashed line).

The corresponding pore space functions, as well as the Gauss fit for the retrospective correction, are shown in Figure 8b. The Gauss corrected (green) and the filter approach (magenta) pore space functions are in good agreement with the expectation and show only slight deviations from intraporal simulation and the ideal cylinder projection. The filter approach again shows a slight baseline, which decays in the outer parts of the local space.

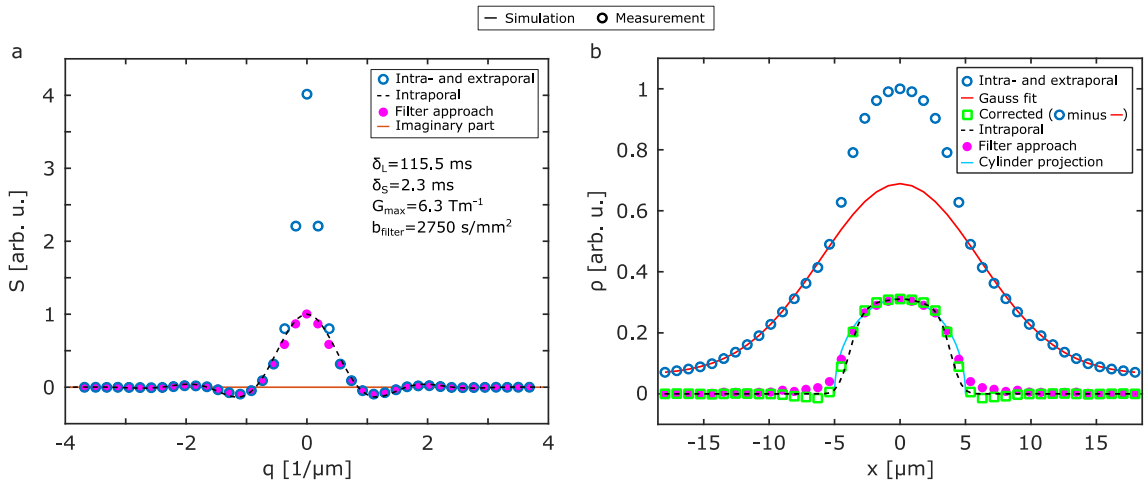

Figure 8: Experimental demonstration of the Gaussian fit and the filter approach. (a) Simulated ground truth data with only intraporal contribution (black), as well as measured signals with (magenta) and without (blue) the filter for glass capillaries with a diameter of 10 µm. The resulting projection of the pore space functions and the Gauss fit (red) are shown in (b). The corrected pore space function by the Gauss fit in $x$ space as well as the filter approach are in good agreement with the expected pore space function.

### 5. Discussion

Characterizing the cell shape and size by NMR could be of high interest for the characterization of biological tissues and thus pathologies. Diffusion pore imaging has the potential to lead to an in vivo, non-invasive histology-like insight into cell systems. Usage is currently limited to high gradient amplitude spectrometers and small animal scanners. In the future it might be possible to achieve local gradient fields by using purposely designed gradient coils [43-45], which might enable the use of diffusion pore imaging in the human context. The use of the Diff30 probe with ultra-high gradient amplitude for the spectrometer allows in principle measurements with a resolution in the range of cellular sizes. Our proposed methods form a step toward the possibility to acquire the shape and size of pores in their natural surroundings. While the measurements times amounted to 14 hours for each of the measurements, it hast to kept in mind that the total water content inside the pores only amounted to about 10 µL and the 5 molar NaCl solution further shielded the signal due to the large conductivity. Larger capillary phantoms could not be constructed because of to the limited space inside a 5mm diameter NMR-tube in the x-y-plane. For actual cell samples or different phantoms, measurement times could be significantly reduced. However, for the current CPMG implementation a minimum set of 16 averages is always necessary to satisfy the minimum necessary phase cycling of the 180° pulses in order to suppress unwanted echo pathways for measurements with small $q$ values. However, depending on the sample, using CPMG approaches might not be necessary, since the mainly susceptibility-induced field inhomogeneities generated by the capillaries orientated perpendicular to the $B_0$ field result in the problematic short $T_2^*$ times observed in our experiments.

## 5.1 Simulations

For both the equilateral triangles and the cylinders, the Gauss correction method and the filter approach yield reliable results, which are in good agreement with the intraporal simulations. The extrapolation approach, which can only be used for the symmetric cylinders, is also in good agreement with the expectation. This approach shows, however, undershoots close to the boundary of the cell. These undershoots in the pore space function interestingly have no clearly visible influence on the reconstructed pore image which may also be due to scaling of the color map.

The difference in contrast between the triangles and cylinders for the images with both intra- and extraporal signal contribution can be explained by the different packing densities of the simulated geometries. The packing density for the equilateral triangles is quite low ($f_e = 0.6$) due to the distribution of triangles on a regular grid which contains large parts of empty space, and therefore the difference in Figure 4(e-g) is much more pronounced as for the cylinders in Figure 5(c-f), where the extraporal volume fraction was $f_e = 0.3$. For even higher packing densities (i.e. lower $f_e$) which are to be expected in vivo [46], the influence of the extracellular signal components will be lower, so that an even a better performance of the correction methods can be assumed. It has to be noted however, the effectiveness of correction and suppression methods is also determined by the effective intra- and extracellular diffusion coefficients. For a high packaging density the effective extracellular diffusion coefficient might be lower which may reduce the efficacy of filter gradient approaches.

## 5.2 Measurements

While measurements with gradient amplitudes of up to $G_{\max} = 6.3$ T/m for the short gradient pulse were used in this study, dealing with gradient pulses of this amplitude is quite challenging. In order to compensate for small gradient imperfections (e.g. eddy currents and limited gradient raster resolution, thermal stability and DC offsets of the gradient power amplifier), it was necessary to adjust the length of the short gradient pulse for each $q$ value individually in order to fulfill the rephrasing condition and furthermore to adjust the pre-emphasis for the gradient amplifier before every measurement, which could be reduced in future using gradient amplifiers with e.g. better thermal compensation and DC offset nulling.

While the extrapolation approach yields good results for the simulation as well as the measurement data, it must be kept in mind, that the method is not generally applicable and will be limited to geometries, sample and measurement parameters for which a range of $q$ values exists where necessary conditions of already suppressed signal of the extraporal space is fulfilled. At the same time the Gaussian phase approximation has still to be valid for the intraporal compartment. For the example of a cylindrical geometry shown here using simulations and measurements, the second condition is only in part fulfilled, since the signal starts to deviate from the Gaussian phase approximation significantly near the first zero crossing but can still be reasonably modeled by the quadratic fitting function used here. Therefore, this method is certainly dependent on the shape of the form factor and will not be applicable to arbitrary geometries without special consideration of these requirements.

The proposed correction method using a Gauss fit in $x$ space is not impeded by these limitations, but solely requires the validity of the Gaussian phase approximation for the extraporal fluid and different effective diffusion coefficients of the compartments in the long-time limit. However, this method relies on the availability of an estimate of the expected pore size to choose the density of $q$ space sampling appropriately to obtain the necessary field of view and to determine the region used for the Gauss fit. The outer parts of the pore space function in $x$ space are quite prone to artifacts at low $q$ values, which was observed in some of our measurements preventing a successful Gaussian fit. Especially, insufficient spoiler gradients and phase cycling may affect the signal quality for low $q$ values.

Therefore, for actual measurements of unknown pores, it might be beneficial to use the filter implementation of the long-narrow approach, as this method does not need any further assumptions or knowledge of the shape and size of the pores in the volume of interest, but only relies on the difference of effective diffusion coefficients in the long-time limit. However, the filter decreases the measured intraporal signal due to $T_2$ decay, additional RF pulses and increased diffusion weighting if the long time-limit is not fully reached within the duration of the filter gradient $\delta_\mathrm{F}$. For $\delta_\mathrm{F} = 42$ ms, which was used in this study, the signal loss due to filter diffusion weighting is not substantial when looking at pores with a diameter of 10 μm.

The proposed methods prove the feasibility of diffusion pore imaging in the presence of extraporal water. With special hardware on preclinical scanners or local gradient coils [43-45] it might be possible to employ this technique to in vivo measurements or at least to actual cell samples in the future. In comparison to other pore imaging measurements so far [13, 33-35, 38, 47], the correction approaches proposed here extend these methods which may enhance applicability to more realistic systems and thus bring pore imaging closer to an actual usage.

## 6. Conclusion

Diffusion pore imaging in the presence of extraporal signal components is possible using the proposed correction methods; furthermore, using the filter gradient approach, pore images can be obtained without any correction by postprocessing steps. For very high packing density, possibly no correction whatsoever is necessary, since the extraporal signal is suppressed at relatively low $q$ values and signal deviations near $q = 0$ mainly lead to a global signal level shift in $x$ space. While general limitations such as the high necessary gradient amplitude are still a concern for the application of diffusion pore imaging for in vivo measurements, cell samples or small animals on preclinical systems will be the next step to be investigated using the correction methods presented here. Furthermore, consideration of the effects on cell membrane permeability in detail will be necessary. Regarding human in vivo applications, especially local gradient coils might open a perspective due the available high gradient amplitudes, which may enable pore imaging in vivo allowing for non-invasive measurements of cell size distributions in the human body, potentially opening a new window for further insight into pathologies.

# Appendix

Table A1: Phase-cycling scheme used for the measurements with the CPMG-long-narrow implementation in this study

| $ph_1$ [°]  | 0   | 90  | 180 | 270 | 0   | 90  | 180 | 270 | 0   | 90  | 180 | 270 | 0   | 90  | 180 | 270 |
| $ph_2$ [°]  | 90  | 0   | 90  | 0   | 270 | 180 | 270 | 180 | 90  | 0   | 90  | 0   | 270 | 180 | 270 | 180 |
| $ph_3$ [°]  | 90  | 0   | 90  | 0   | 270 | 180 | 270 | 180 | 270 | 180 | 270 | 180 | 90  | 0   | 90  | 0   |
| $ph_{31}$ [°] | 180 | 270 | 0   | 90  | 180 | 270 | 0   | 90  | 180 | 270 | 0   | 90  | 180 | 270 | 0   | 90  |